\documentclass[twocolumn,showpacs,preprintnumbers,amsmath,amssymb]{revtex4}

\usepackage{graphicx}% Include figure files
\usepackage{dcolumn}% Align table columns on decimal point
\usepackage{bm}% bold math
\usepackage[T2A]{fontenc}
\usepackage[cp1251]{inputenc}
\usepackage[dvips]{color}

\begin{document}

\title{EIT-based Vector Magnetometry
in Linearly Polarized Light}

\author{V. I.~Yudin$^{1,2,3}$, A. V.~Taichenachev$^{1,2}$, Y. O.~Dudin$^{4}$, V. L.~Velichansky$^{5,6}$, A. S.~Zibrov$^{7}$, S. A.~Zibrov$^{6}$ \\
 $^{1}${\em Institute of Laser Physics, Siberian Branch of RAS,  Novosibirsk, 630090, Russia}\\
 $^{2}${\em Novosibirsk State University, Novosibirsk, 630090, Russia}\\
 $^{3}${\em Novosibirsk State Technical University, Novosibirsk, 630092, Russia}\\
 $^{4}${\em School of Physics, Georgia Institute of Technology, Atlanta, Georgia, 30332-0430, USA}\\
 $^{5}${\em Moscow State Engineering and Physics Institute, Moscow, 115409, Russia}\\
 $^{6}${\em Lebedev Physical Institute, RAS, Moscow, 117924, Russia}\\
 $^{7}${\em Physics Department, Harvard University, Cambridge, MA, 02138, USA}
 }
\date{\today}

\begin{abstract}
We develop a generalized principle of EIT vector magnetometry based on high-contrast EIT-resonances and the symmetry of atom-light interaction in the linearly polarized bichromatic fields. Operation of such vector magnetometer on the $D_{1}$ line of $^{87}$Rb has been demonstrated. The proposed compass-magnetometer has an increased immunity to shifts produced by quadratic Zeeman and ac-Stark effects, as well as by atom-buffer gas and atom-atom collisions. In our proof-of-principle experiment the detected angular sensitivity to magnetic field orientation is $10^{-3}$deg$/$Hz$^{1/2}$, which is limited by laser intensity fluctuations, light polarization quality, and magnitude of the magnetic field.
\end{abstract}


\maketitle

\section{Introduction}

The pure quantum state is a  basic concept of quantum physics, which plays a key role in various applications, such as magnetometry, frequency standards, laser cooling,  quantum information science, nonlinear optics, and ``slow'' and ``fast'' light experiments. The effect of electromagnetically induced transparency (EIT) \cite{Alzetta,Arimondo,Harris,Fleisch} has been successfully employed in all these applications.

The idea of EIT scalar magnetometer has been suggested in
\cite{Scully}. The steep dispersion of  EIT media  promises a
dramatic improvement of the scalar magnetometer sensitivity. Since
then different schemes for EIT magnetometry have been considered. Among
them are schemes based on the nonlinear Faraday effect in a manifold of
a single ground state \cite{Budker2,Novikova, Stahler} and a scheme
in which the frequency shift of Zeeman sublevels of both ground
states is detected \cite{Wynands}. The sensitivity of EIT
magnetometers is in the same range as magnetometers using optical
pumping \cite{BudBud,Bud}. The recent modification of
optically pumped  magnetometers with
suppressed spin-exchange broadening (so-called SERF-magnetometer) drastically improves
sensitivity by a factor of $10^3$. It overcomes the sensitivity of
SQUID magnetometers ($10^{-15}$~T$/$Hz$^{1/2}$) \cite{Romalis}.
Unfortunately, SERF-magnetometers work in small fields that are less than 0.1 $\mu$T, which is
significantly weaker than geomagnetic field.

For many applications it is preferable to know not only the scalar
-- but also the direction of the magnetic field. To achieve this,
individual coils  are installed  for each of the $X$, $Y $, and $Z$
axes in a scalar magnetometer. The coils are used to induce small
modulations of the magnetic field along each axes, which gives the
information about $B_x$ and $B_y$ field components
\cite{Fairweather,Gravrand,Alexandrov2}. This allows the orientation
of the vector $\textbf{B}$ to be reconstructed. The first
schemes of EIT vector magnetometer have been proposed in
\cite{Wynands2,Lee}. However, the angular accuracy of these
magnetometers strongly depends on mathematical models (describing
the atom-field interaction and light field propagation) used to
extract the magnetic field direction from experimental signals. The
reviews of existing all-optical magnetometers were published in
\cite{Romalis2,Alexandrov}.

In the present paper we show that employing the unique features of
high-contrast EIT resonances on the $D_1$ line of $^{87}$Rb allows
us to find new approaches to the atomic vector magnetometery and to
model a relevant device in which the scalar and vector properties of
magnetic field can be measured separately or simultaneously. Our approach does not require the mathematical models to reconstruct  a three-dimensional orientation of the magnetic field.

\section{General description of the problem}

EIT phenomenon is closely connected to the so-called coherent population trapping (CPT) \cite{Alzetta,Arimondo} in which the atom-field interaction $-(\widehat{\bf d}{\bf E})$ of the pure quantum state $|dark\rangle$ is zero:
\begin{equation}\label{dark}
   -(\widehat{\bf d}{{\bf E}})|dark\rangle = 0\,.
\end{equation}
This state is a special coherent superposition of the ground state
Zeeman sublevels that neither absorbs nor emits light. Dark
states lead to the highest contrast of  EIT
resonances. Thus, the preparation of pure states is
crucial for any of the above mentioned applications.

\begin{figure}[h]
  \includegraphics[width=8.5cm]{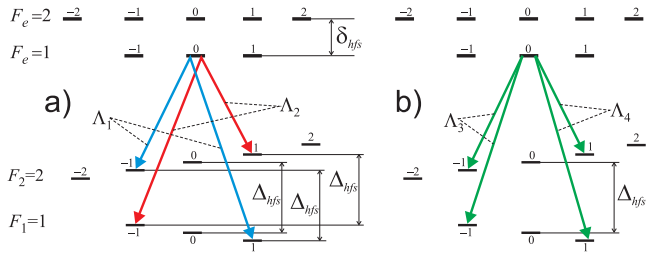}
\caption{
Pure $\Lambda$-systems at the $D_1$ line of $^{87}$Rb: non-sensitive~(a) and sensitive~(b) to magnetic field.
Here we do not show Zeeman shifts for upper hyperfine levels with $F_e$=1,2.}\label{levels}
\end{figure}

\begin{figure}[]
  \includegraphics[width=2.5in]{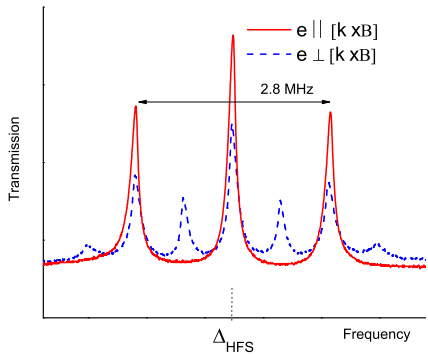}\hfill
  \caption{EIT resonance transmission:\\
(solid line) -- the case $\mathbf{e}$$\parallel$$\mathbf{n}$. The
central resonance corresponds to the $\Lambda_1$ and $\Lambda_2$
schemes (Fig.~\ref{levels}a). This resonance has
120~kHz width and  $\sim60$$\%$ transmission.\\
(dashed line) -- the case $\mathbf{e}$$\perp$$\mathbf{n}$. Magnetic
field has magnitude 1~G, the angle between $\mathbf{B}$ and
$\mathbf{k}$ equals 20$^{\circ}$.} \label{general_view_20}
\end{figure}

The generalized problem of the production of pure quantum states by
bichromatic elliptically polarized field was solved in \cite{Y}. In
\cite{icono05,lin||lin JETP,Serezha} it was theoretically and
experimentally demonstrated that the $D_1$ line of $^{87}$Rb has
unique level structure for the production of pure dark states using
bichromatic linearly polarized light (so-called lin$||$lin field),
where the resonant interaction occurs via the upper energy level
$F_e$=1. There are two pairs of dark states, where  each dark state
corresponds to the separate $\Lambda$-scheme (see
Fig.~\ref{levels}). One pair corresponds to $\Lambda_1$ and
$\Lambda_2$ schemes in Fig. 1a and involves the following two-photon
transitions: $|F_{1}=1,m=-1 \rangle \leftrightarrow |F_2=2,
m=+1\rangle$ and $|F_{1}=1, m=+1 \rangle\leftrightarrow |F_2=2,
m=-1\rangle$. In our experiments the EIT resonances of these pairs
have a high contrast (50{\%}) and transmission (60{\%}) (solid line
in Fig.~\ref{general_view_20}). Both $\Lambda_1$ and $\Lambda_2$
transitions contribute to EIT resonance (the dependence of
transmission on the difference of the two optical frequencies) that
is attractive for applications in chip-size atomic clocks (CSAC)
since it provides high contrast and smaller (by factor 1.33)
quadratic dependence on the magnetic field compared to the regular
atomic clock transition $|F_{1}=1, m=0 \rangle\leftrightarrow
|F_2=2, m=0\rangle$ \cite{Evelina,Novikova2,Mikhailov2}. Note that
the shifts of zero magnetic sublevels and the frequency of 0-0
transition do not depend linearly on magnetic field, while sublevels
with $m=\pm 1$ do. The electron $g$-factors of the ground states
$F_{1,2}$ have the same  magnitude but \textit{opposite sings} (see
Fig.~\ref{levels}). As a result, the residual linear shifts (due to a
nuclear contribution) of the $\Lambda_1$ and $\Lambda_2$ transitions
are 250 times smaller than the shifts of individual magnetic
sublevels  $m=\pm 1$ ($\approx\pm$28~Hz$/$$\mu$T instead of
$\approx\pm$7~kHz$/$$\mu$T). However, these residual shifts are manifested only in a small broadening of the resonance lineshape, while the center of the resulting $\Lambda_{1,2}$-resonance has a zero linear sensitivity to the magnetic field (due to the symmetry of $\Lambda_1$ and $\Lambda_2$ systems for the lin$||$lin light) \cite{icono05,lin||lin JETP}.

The other pair of $\Lambda$-schemes ($\Lambda_3$ and $\Lambda_4$ in
Fig.~\ref{levels}b) gives the two-photon transitions: $|F_2=2;
m=-1\rangle\leftrightarrow|F_1=1; m=-1\rangle$ and $|F_2=2;
m=+1\rangle\leftrightarrow|F_1=1; m=+1\rangle$ that strongly depend
on magnetic field and can be used for measurment of the magnetic
field magnitude, as it was noted in \cite{lin||lin JETP}.

To produce quantum dark states (\ref{dark}) for the $D_1$ line of
$^{87}$Rb, we use (in conformity with \cite{icono05,lin||lin JETP})
a linearly polarized bichromatic running wave ${\bf E}({\bf r},t)$
with close frequencies $\omega_1$ and $\omega_2$ and wavevector
$\mathbf{k}$ (i.e. lin$||$lin configuration):
\begin{equation}\label{1}
{\bf E}({\bf r},t)=(E_1e^{-i\omega_1t}+ E_2e^{-i\omega_2t})e^{i{\bf kr}}{\bf e}+c.c.\,,
\end{equation}
where ${\bf e}$ is a unit vector of the linear polarization, and $E_{1,2}$ are the scalar amplitudes of the corresponding frequency components. The interaction occurs in the presence of the static magnetic field ${\bf B}$. If the $z$-axis is directed along the vector ${\bf B}$, the vector ${\bf e}$ can be expressed in a spherical basis \{${\bf
e}_0$=${\bf e}_z$, ${\bf e}_{\pm 1}$$=$$\mp$$({\bf e}_x$$\pm$$i{\bf e}_y)$/$\sqrt{2}$\}:
\begin{eqnarray}\label{e}
{\bf e}&=&\sum_{q=0,\pm 1}e^{(q)}{\bf e}_q=\cos{\theta}\,{\bf e}_0-\frac{\sin{\theta}}{\sqrt{2}}(\bf{e}_{+1}-\bf{e}_{-1})\,,
\end{eqnarray}
where $\theta$ is the angle between vectors ${\bf B}$ and ${\bf e}$; $e^{(q)}$ are the contravariant components of the vector ${\bf e}$. Note that for linear polarization its  circular components ($\sigma_{\pm}$) are always equal:
\begin{equation}\label{e_pm}
|e^{(+1)}|=|e^{(-1)}|=|\sin{\theta}|/\sqrt{2}\,.
\end{equation}
As it will be shown below, the symmetry (\ref{e_pm}) is one of
principal points of EIT magnetometry in a linear polarized
field.

In the resonant approximation we assume that the frequency component
$\omega_j$ ($j$=1,2) excites atoms only from the hyper-fine ground
level $F_j$ (Fig.~\ref{levels}). From here on, we use the
interaction representation
$$e^{-i{\cal E}^{}_{Fm}t/\hbar}|F,m\rangle \to|F,m\rangle \,,$$
where ${\cal E}^{}_{Fm}$ is the energy of the level $|F,m\rangle$ in which the Zeeman shift is included. The operator of an atom-field  interaction $-(\widehat{\bf d}{\bf E})=\widehat{V}+\widehat{V}^{\dagger}$ under the resonant approximation takes the form:
\begin{eqnarray}\label{V}
\widehat{V}&=&e^{i{\bf kr}}\sum_{q=0,\pm 1} e^{(q)}\times  \\
&&{}\biggl[ E_1\sum_{F_e,\mu,m^{}_1}d^{}_{F_e F_1}e^{-i\delta^{(1)}_{\mu m_1}t}C^{F_e \mu}_{F^{}_1 m^{}_1,1q}
|F_e,\mu\rangle\langle F^{}_1,m^{}_1|+ \nonumber \\
&& E_2\sum_{F_e,\mu,m^{}_2}d^{}_{F_e F_2}e^{-i \delta^{(2)}_{\mu m_2}t}C^{F_e\mu}_{F^{}_2 m^{}_2,1q}
|F_e,\mu\rangle \langle F^{}_2,m^{}_2|\biggr].\nonumber
\end{eqnarray}
Here $d^{}_{F_eF_1}$ and $d^{}_{F_eF_2}$ denote reduced matrix elements of  corresponding optical transitions $F_1\to F_e$ and $F_2\to F_e$, $C^{F_em_e}_{F^{}_jm^{}_j,1q}$ are Clebsch-Gordan coefficients, and
$\delta^{(j)}_{\mu m^{}_j}=\omega_j-({\cal E}^{}_{F_e\mu} -{\cal E}^{}_{F^{}_jm^{}_j})/\hbar$ for $j=1,2$ are corresponding one-photon detunings.

For alkaline atoms with nuclear spin $I_n$ we have
$F_1$$=$$(I_n$$-$$1/2)$ and $F_2$$=$$(I_n$$+$$1/2)$. The
corresponding electron ground-state Land$\acute{\mbox{e}}$ factors
have the same absolute value but opposite signs:
$g$=$-g^{}_{F_1}$=$g^{}_{F_2}$=($I_n$$+$$1/2)^{-1}$ (for $^{87}$Rb
$g$=$1/2$). Then in the linear approximation of the dependence on
the magnetic field ${\bf B}$ and neglecting the nuclear magneton
contribution it is easy to count the number of splitted two-photon
resonances. For arbitrary directed ${\bf B}$ there are (4$I_n+$1)
two-photon resonances in transmissions versus Raman detuning
$\delta_R$=($\omega_1$$-$$\omega_2$$-$$\Delta_{hfs}$) dependencies
centered at the points $\delta_R$=$l$$g$$\mu^{}_B$$|{\bf
B}|$/$\hbar$ ($l$=$-2I_n$,...,$2I_n$), where $\mu^{}_B$ is Bohr
magneton. For example, in $^{87}$Rb ($I_n$=3/2) we have seven
two-photon resonances (the blue dashed line in
Fig.~\ref{general_view_20}). In the particular case of
$\mathbf{B}$$\perp$$\mathbf{e}$, the number of two-photon resonances
equals to $2I_n = 3$ (the red solid line in
Fig.~\ref{general_view_20}).

\section{EIT-based 3D compass}

First we examine in detail the central resonance (near
$\delta_R$=0). It will be shown below that this resonance can be
used for the vector magnetometer due to the strong dependence of
transmission on the mutual orientation of vectors ${\bf e}$ and
${\bf B}$ (i.e. on the angle $\theta$ in Eq.(\ref{e})). The
following two transitions take place in formation of the central
two-photon resonance: $|F_1$,{\em
m}=$-1\rangle$$\leftrightarrow$$|F_2$,{\em m}=$+1\rangle$ and
$|F_1$,{\em m}=$+1\rangle$$\leftrightarrow$$|F_2$,{\em
m}=$-1\rangle$, for which the energy difference equals
$\hbar$$\Delta_{hfs}$ (Fig.~\ref{levels}a). The third two-photon
transition $|F_1$,{\em m}=$0\rangle$$\leftrightarrow$$|F_2$,{\em
m}=$0\rangle$ (between magnetically insensitive sublevels) is
strongly suppressed due to further destructive interference of
contributions from the opposite circular components $\sigma_{\pm}$.

In the case of a resolved  upper hyper-fine structure ($F_e$=1,2 in
the Fig.~\ref{levels}), the two-photon resonance can be excited via
separate level. Further we assume that the frequency components
(\ref{1}) are at the resonance with a single hyper-fine level
$F_e$=1 (Fig.~\ref{levels}). Now let us consider a special case
where the vectors ${\bf e}$ and ${\bf B}$ are mutually orthogonal
($\theta$=$\,\pi/2$), and, therefore, only two equal circular
components ${\bf e}=-({\bf e}_{+1}-{\bf e}_{-1})/\sqrt{2}$ occur in
the decomposition (\ref{e}). It is seen from Fig.~\ref{levels}a that
there is a two-photon resonance formed via pure $\Lambda_1$-scheme
with Zeeman sublevels $|F_1$=1,{\em m}=$+1\rangle$ and $|F_2$=2,{\em
m}=$-1\rangle$. Similarly, the $\Lambda_2$-scheme is realized with
the other sublevel pairs $|F_1$=1,{\em m}=$-1\rangle$ and
$|F_2$=2,{\em m}=$+1\rangle$. Both of these
$\Lambda^{}_{1,2}$-schemes are formed via the same common upper
sublevel $|F_e$=1,{\em m}=$0\rangle$. As was mentioned before, the
frequencies of these two-photon resonances are equal (neglecting the
nuclear magneton contribution) to the frequency of the (0-0)
resonance between sublevels $|F_1$=1,{\em m}=$0\rangle$ and
$|F_2$=2,{\em m}=$0\rangle$.

The uniqueness of the situation arises from the overlapping the two
($|dark\rangle_{\Lambda^{}_{1}}$ and
$|dark\rangle_{\Lambda^{}_{2}}$) dark states, which occur at the
two-photon resonance, $\delta_R=0$. These states satisfy the
equation (\ref{dark}) and have the following forms:
\begin{equation}\label{D_pm}
|dark\rangle_{\Lambda^{}_{1,2}}=\frac{\sqrt{3}E_{2}|F_1,m=\pm
1\rangle \mp E_{1}|F_2,m=\mp 1\rangle}{\sqrt{|E_{1}|^2+3|E_{2}|^2}}.
\end{equation}
The presence of such dark states in the  ${\bf e}$$\perp$${\bf B}$
case leads to a high contrast of the central dark resonance near
$(\omega_1-\omega_2)$=$\Delta_{hfs}$ (i.e. $\delta_R$$\approx$$\,0$).
This fact was predicted and experimentally demonstrated in \cite{icono05,lin||lin JETP}.

In the general case of $\theta$$\neq$$\pi/2$ (that is
$\cos(\theta)$$\neq$0) there are no  pure $\Lambda$-schemes due to
the $\pi$-polarized (along ${\bf B}$) component in decomposition
(\ref{e}). It leads to a smaller amplitude and contrast of the
central two-photon $\Lambda_{1,2}$-resonance in comparison to the
case of $\theta$=$\pi/2$. This fact will be used as a basis for
determination of the magnetic field orientation (i.e. compass) in
our approach.

\begin{figure}[t]
\centerline{\scalebox{0.35}{\includegraphics{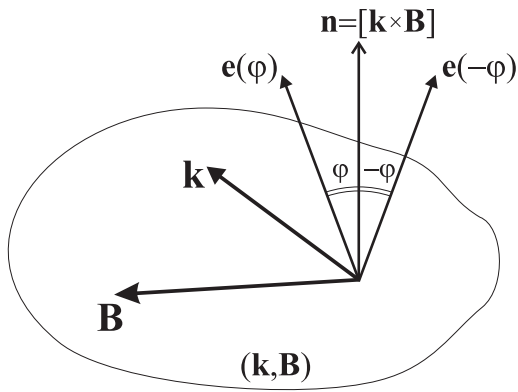}}}\caption{Orientation of the magnetic field $\textbf{B}$, wavevector  $\textbf{k}$ and polarization  $\textbf{e}$ of the optical field.
}\label{plane}
\end{figure}

The basic idea of our method can be explained in the following way.
Assume that the wavevector ${\bf k}$ and the vector ${\bf B}$ have
an arbitrary mutual orientation. We will use the amplitude of the
central resonance (absorption, transmission or fluorescence) as the
measured quantity (Fig.~\ref{general_view_20}). There are two cases,
where \textbf{e} and \textbf{B} are orthogonal to each other. More
precisely, these situations arise if ${\bf e}$$||$${\bf n}$, where
${\bf n}$=$[{\bf k}\times{\bf B}]$ (Fig.~\ref{plane}). These cases
correspond to the dark states (\ref{D_pm}), which lead to the
maximal amplitude and contrast of the central resonance (as
explained above).

Consider the dependence of the dark resonance amplitude, which is
obtained by rotating the polarization vector ${\bf e}$ around fixed
wavevector ${\bf k}$. This dependence can be presented as a function
$A_{\bf k}$($\varphi$), where $\varphi$ is the angle between the
vectors ${\bf e}$ and ${\bf n}$ (Fig.~\ref{plane}). Even the
qualitative analysis, provided above, leads to the conclusion  that
the function $A_{\bf k}$($\varphi$) reaches its maximum at
$\varphi$=0,$\,\pi$, i.e. when ${\bf e}$$\perp$${\bf B}$.

The essence of the measuring procedure could be represented by the
following algorithm. At first, for a chosen vector ${\bf k}$=${\bf
k}_1$, we get the $A_{{\bf k}_1}$($\varphi$) dependence by rotating
the polarization vector ${\bf e}$ around wavevector ${\bf k}_1$. The
maximum of this dependence corresponds to the direction of the
vector ${\bf n}$=$[{\bf k}_1\times{\bf B}]$, which gives us  the
equation for the plane (${\bf k}_1,{\bf B}$) formed by the vectors
${\bf k}_1$ and ${\bf B}$. Repeating the same procedure for another
orientation of the wavevector ${\bf k}$=${\bf k}_2$ (for example,
${\bf k}_2$$\perp$${\bf k}_1$) provides the equation for the plane
(${\bf k}_2,{\bf B}$). The intersection line of the two planes
(${\bf k}_1,{\bf B}$) and (${\bf k}_2,{\bf B}$) gives the 3D
orientation of the vector ${\bf B}$ with an uncertainty of the sign.

The basic principle of our method is quite universal and does not
depend on different experimental parameters (such as the $|E_1/E_2|$
ratio, one-photon detuning, relaxation constants, atom-atom
collisions, nuclear magnetic momentum, and so on). This can be seen
from the general symmetry of the problem. Indeed, suppose we have an
arbitrary polychromatic wave propagating along a direction ${\bf k}$
and having the same linear polarization ${\bf e}$ for all frequency
components. Also we assume that the  atomic medium is isotropic in
the absence of the light field. We determine the signal $S$({\bf
e},{\bf B}) as a scalar value that depends on the mutual orientation
of the vectors \textbf{e} and \textbf{B}. In the sense of this
definition, $S$({\bf e},{\bf B}) could be the transmission,
absorption, or fluorescence. A general analysis of the Bloch
equations gives the following relationship:
\begin{equation}\label{B-B}
S({\bf e},{\bf B})=S({\bf e},-{\bf B})=S(-{\bf e},{\bf B})\,.
\end{equation}
The left equality comes from the symmetry of Clebsch-Gordan
coefficients $|C^{F'm'}_{Fm,1q}|=|C^{F'-m'}_{F-m,1-q}|$ and the
equality of the circular components (\ref{e_pm}) in an arbitrary
coordinate system. The right equality in (\ref{B-B}) arises due to
an independency of the $S$({\bf e},{\bf B}) on field phase (the
transmission and absorption depend on the $|{\bf E}|^2$).

Consider a configuration shown in  Fig.~\ref{plane}, where the light
field has {\bf e}($\varphi$ polarization ). Let us perform a
mathematical reflection in relation to the plane (${\bf k},{\bf
B}$). This leads to the substitution of the polarization vector {\bf
e}($\varphi$)$\to$($-{\bf e}$($-\varphi$)), but for the {\em
pseudovector} of the magnetic field it leads to  ${\bf
B}$$\to$($-{\bf B}$). It is known, that  the mathematical reflection
does not affect a scalar signal, i.e. another relationship is
obtained:
\begin{equation}\label{e-e}
S({\bf e}(\varphi),{\bf B})=S(-{\bf e}(-\varphi),-{\bf B})\,.
\end{equation}
By combining (\ref{B-B}) and (\ref{e-e}) we finally achieve
\begin{equation}\label{phi-phi}
S({\bf e}(\varphi),{\bf B})=S({\bf e}(-\varphi),{\bf B})\,,
\end{equation}
i.e. the scalar signal is an even function of the angle $\varphi$.
It means that the points $\varphi$=0,$\,\pi$ (i.e. ${\bf
e}$$\perp$${\bf B}$) correspond to the local extremum (maximum or
minimum) of the $S({\bf e}(\varphi),{\bf B})$ dependence, which is
obtained by rotating the polarization vector ${\bf e}$ around
wavevector ${\bf k}$. Similar symmetry consideration shows
that there are two other extremuma when the vector ${\bf e}$ lies in
the plane (${\bf k},{\bf B}$), i.e., at $\varphi$=$\pm\pi/2$. Note,
that the derived results remain valid in the case of light wave
propagation (including the nonlinear effects in an optically thick
medium). In this case the used above vector {\bf e}($\varphi$)
corresponds to the initial linear polarization before atomic medium
(cell).

Thus, we have shown that the described principle of EIT vector
magnetometery is valid, in essence, for arbitrary atoms, lines
($D_1$ or $D_2$), and arbitrary spectral composition of linearly polarized
wave (including a monochromatic wave). However, the dependence
$A_{\bf k}$($\varphi$) for the  central resonance excited by a
bichromatic field at the $^{87}$Rb $D_1$ line is the best choice for
the demonstration of the compass principle because of the
significant signal$/$noise ratio and transmission.

EIT vector magnetometry in a circularly polarized light has been discussed  in \cite{Wynands2,Lee}. In those schemes
the mathematical models (density matrix and Maxwell equations) are required to reconstruct the vector of the magnetic field from experimental signals. Meanwhile, any model is sensitive to the ultimate knowledge of involved parameters and processes, such as: light intensities, one-photon detunings, light beam profile, atomic density, atomic diffusion motion in buffer gas, collision processes (depolarization, broadening, shifts), and so on. This may limit and sufficiently decrease the achievable angular accuracy (to the level of 1-10~deg) of the vector magnetometer. In contrast, our 3D compass does not require the use of mathematical models, because the extremum of the angle dependence $A_{\bf k}$($\varphi$) at the points $\varphi$=0,$\,\pi$ is an inherent feature.

\section{EIT scalar magnetometer}

As was shown above, rotating the linear polarization ${\bf e}$
around the wavevector ${\bf k}$ and analyzing the corresponding
dependence of amplitude $A_{\bf k}$($\varphi$) of the central dark
resonance, we always can find the condition ${\bf e}$$\perp$${\bf
B}$. In this section we will consider two end magnetically sensitive
resonances (Fig.~\ref{general_view_20}, red solid line), which are
connected with $\Lambda_{3,4}$-systems shown in Fig.~\ref{levels}b
(i.e. with two-photon transitions $(-1)$$\leftrightarrow$$(-1)$ and
$(+1)$$\leftrightarrow$$(+1)$). In the ${\bf e}$$\perp$${\bf B}$
case, the amplitudes of these resonances attain maximum too, because
at the exact two-photon resonance (i.e. $\delta_R$=$\pm
2g$$\mu^{}_B$$|{\bf B}|$/$\hbar$) there are the following two dark
states:
\begin{equation}\label{D_34}
|dark\rangle_{\Lambda^{}_{3,4}}=\frac{\sqrt{3}E_{2}|F_1,m=\mp
1\rangle \pm E_{1}|F_2,m=\mp 1\rangle}{\sqrt{|E_{1}|^2+3|E_{2}|^2}}.
\end{equation}
We can determine the value $|{\bf B}|$ by measuring the distance between these resonances $|\Delta_{\pm}|$. In the linear approximation for $|{\bf B}|$ we apply the formula $|\Delta_{\pm}|$=$\gamma'$$|{\bf B}|$, where $\gamma'$=$2(g^{}_{F_2}-g^{}_{F_1})$$\mu^{}_B$$/$$\hbar$ is an effective gyromagnetic ratio. Due to the effect of the nuclear magneton for $^{87}$Rb \cite{Arimondo77,steck} we should use the following values for $g$-factors: $g^{}_{F_1}$=$-$0.501827 and $g^{}_{F_2}$=0.499836. Thus, in our case $\gamma'$=2.803905$\times$10$^{10}$~Hz$/$T.

Taking into account the symmetry of the atom-light interactions in the linear polarization one can predict some important properties of such magnetometry scheme. Indeed, this frequency-differential magnetometer is immune to:\\
{\bf (I)}   the collisional shift arising due to interactions with an isotropic buffer gas;\\
{\bf (II)}   the quadratic Zeeman shift of  magnetic sublevels;\\
{\bf (III)}  the shift arising from atom-atom interactions (including spin-exchange)
between atoms (here between $^{87}$Rb atoms);\\
{\bf (IV)} the ac-Stark shift.

The property {\bf (I)} is a result of the  equality  of the
collisional shifts of all Zeeman sublevels $|F,m\rangle$ (for a
given $F$) in an isotropic buffer gas. The property {\bf (II)}  is
also quite obvious considering that each even (on $|{\bf B}|$) power
of the Zeeman shift has an equal value and sign for the
$m$$\leftrightarrow$$m$ and ($-m$)$\leftrightarrow$($-m$)
transitions. This feature is valid for any atom (i.e. not only for
$^{87}$Rb) and line ($D_1$ and $D_2$).

The property {\bf (III)} is a result of the interaction of atoms
with linear polarized light.  Indeed, let us consider the atomic
density matrix $\widehat{\rho}$, which describes the distribution
among Zeeman sublevels:
\begin{equation}\label{rho}
\widehat{\rho}=\sum_{F',m',F'',m''}\rho^{F'F''}_{m'm''}| F',m'\rangle\langle F'',m''|\,,
\end{equation}
where $\rho^{F'F''}_{m'm''}$ are matrix elements. We denote the
atomic distribution for two-photon resonances
$(+1)$$\leftrightarrow$$(+1)$ and $(-1)$$\leftrightarrow$$(-1)$ (at
the $\delta_R$=$\pm g\mu^{}_B |{\bf B}|$) as
$\rho^{(+)F'F''}_{m'm''}$ and $\rho^{(-)F'F''}_{m'm''}$
respectively. From the general symmetry and neglecting some
insignificant details (for example  a small variation of the
one-photon detuning) we get
$|\rho^{(+)F'F''}_{m'm''}|$=$|\rho^{(-)F'F''}_{-m'-m''}|$.
Obviously, this relationship is not changed by the atom-atom
interactions (including the spin-exchange process). Therefore, the
corresponding collisional frequency shifts have the same magnitude,
i.e. they do not affect the  frequency difference $\Delta_{\pm}$
(while the collisional {\em broadening} of the EIT-resonances will
have an influence). This property gives a significant advantage in
comparison with other schemes of atomic magnetometers, where the
atom-atom interaction is a limiting factor for precise magnetic
field measurements. The property {\bf(III)} supports also the use of
miniature size cells in our EIT magnetometer, because it is possible
to work at high cell temperature to get high atomic density.

Note that the property {\bf (III)} can be extended to an arbitrary element (i.e. not only $^{87}$Rb) and resonance line, when  the magnetometer uses the frequency difference between two-photon resonances $m$$\leftrightarrow$$m$ and $(-m)$$\leftrightarrow$$(-m)$. In general, the angle $\theta$ (between vectors ${\bf B}$ and ${\bf e}$) can be arbitrary.

The property {(\bf IV)} follows from two circumstances. Firstly, the
light shifts of  two-photon resonances (see
Fig.~\ref{general_view_20}, the red solid line) that occur via upper
level $F_e$=1 are absent, because the dark states nullify the
resonant interaction (\ref{dark}). Therefore, these ac-Stark shifts
are small and appear mostly due to the interaction with the
far-off-resonance level $F_e$=2, see Fig.~\ref{levels}. Secondly,
due to the symmetry, these shifts are practically identical and
compensate each other (in the value $\Delta_{\pm}$). There is,
however, a small disbalance caused by Zeeman splitting ($\Delta_Z$).
This splitting leads to a small difference for all one-photon
detunings near the dark resonances $(+1)$$\leftrightarrow$$(+1)$ and
$(-1)$$\leftrightarrow$($-1$). Thus, if the value of the light shift
for extreme resonances is approximately $U$, then the relative shift
can be estimated as $\sim$$|\Delta_Z/\delta_{hfs}|$$|U|$, which
means an additional significant suppression of shift by the factor
$|\Delta_Z/\delta_{hfs}|$$\ll$1. Note, that similar
advantages of the optically pumped balance magnetometer also has
been pointed in \cite{Alexandrov3}.

In our case the magnetometer sensitivity $\delta B$ depends on the
signal-to-noise ratio (S/N) of the Zeeman resonance signal and the
width of the EIT resonance $\Gamma_{\tiny \mbox{FWHM}}$:  $\delta
B={\Delta B}/(S/N)$, where $\Delta B$=$\Gamma_{\tiny
\mbox{FWHM}}$$/$$\gamma'$. Therefore, a high contrast of the
$\Lambda$-resonances, where most of the atoms ((50-70)$\%$) are
accumulated in the dark state \cite{icono05,lin||lin JETP,Serezha},
makes them a perspective competitor for existing all-optical
magnetometers \cite{Alexandrov}. As an example, we estimate the
achievable sensitivity using recently published  data on the
lin$||$lin resonances \cite{Evelina} (authors of \cite{Evelina}
characterized the lin$||$lin resonances as an atomic clock
reference). With a resonance width $\Gamma_{\tiny
\mbox{FWHM}}$=900~Hz and $S/N$=3300~Hz$^{1/2}$, the
sensitivity for the measurable magnetic field is $\delta{B}<
10^{-11}$~T$/$Hz$^{1/2}$, which can be obtained without special
efforts and for very moderate density $10^{10}$~cm$^{-3}$ of
rubidium atoms (50~C$^{\circ}$ and 1.2 Torr $N_2$ pressure in
\cite{Evelina}). To significantly improve the sensitivity one should
increase the number of atoms. In this case the EIT differential
magnetometer will achieve sensitivity at the level $\delta{B}\sim
10^{-13}$-10$^{-14}$~T$/$Hz$^{1/2}$ or better, because we expect to
reach an atom concentration greater than $10^{12}$~cm$^{-3}$ without
serious limitations due to collisional processes (property {\bf
(III)}). The proper choice of buffer gas pressure and the additional
narrowing of EIT resonance in dense media \cite{Mikhailov,Lukin}
also gives some advantages. However, it is worth noting that the
behavior of the coherent effects (EIT) in dense vapor
$>$$10^{12}$~cm$^{-3}$ has not yet been studied in detail, though it
is known that at $10^{14}$~cm$^{-3}$ EIT is still observed
\cite{Matsko2}.

Additionally we note that each of $\Lambda_3$ and $\Lambda_4$ resonances (i.e. $(-1)$$\leftrightarrow$$(-1)$ and $(+1)$$\leftrightarrow$($+1$) two-photon transitions) can be used also in the compass scheme (described in the previous section). But it has some drawbacks in comparison with the compass based on the central resonance (i.e. $\Lambda_{1,2}$). Firstly, the frequency position of each of these resonances depends on the magnetic field. Secondly, their transmission dependence $A_{\bf k}$($\varphi$) vs rotation of the vector $\mathbf{e}(\varphi)$ can have two local maximum. One of them (which always exists) corresponds to the case ${\bf e}$$\perp$${\bf B}$, but the other possible maximum emerges, when the vector ${\bf e}$ lies in the plane (${\bf k}$,${\bf B}$). Such situation leads to an uncertainty in the measurement procedure.

\begin{figure}[]
  \center
  \includegraphics[width=2.38in]{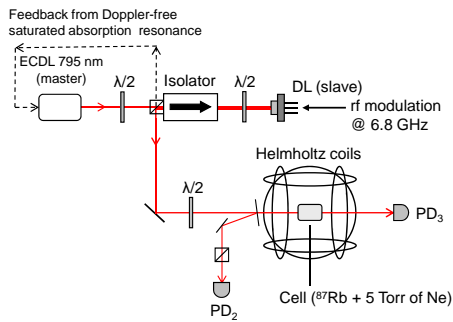}
  \caption{The schematic of the experimental setup.}\label{setup}
\end{figure}

\section{Experiment}

From our point of view the possibility of an EIT-based compass is
the most attractive and unusual part of the suggested ideas.
Therefore, in the experimental part we just concentrate on this
idea. The experimental setup is shown in Fig.~\ref{setup}. The
bichromatic field $\mathbf{E}(\mathbf{r},t)$ is delivered by an
extended cavity (ECDL), frequency is  modulated at
$\Delta_{hfs}$=6.8~GHz and  injected into the slave diode laser DL
\cite{lin||lin JETP}. The experiment is carried out on a Pyrex cell
(40~mm long and 25~mm in diameter) containing isotopically enriched
$^{87}$Rb and 5~Torr neon buffer gas. The cell is placed inside
Helmholtz coils, where the field inhomogeneity  is $\sim$2~mG/cm.
For the experiments reported here the cell temperature is
45$^{\circ}$~C.

The laser frequency is locked to the Doppler-free saturated absorption
resonance. The radiation power at the cell front window is
1.5~mW. To excite the $\Lambda_{1,2}$ scheme the
carrier frequency is tuned to the $F_2 = 2\rightarrow F_e=1$ transition, and the high
frequency side-band is tuned to the $F_1 = 1\rightarrow F_e=1$
transition. The displayed spectra of the EIT resonances are
shown in Fig.~\ref{general_view_20}, where the curves correspond
to the $^{87}$Rb transmission spectra  for the two cases
$\mathbf{e}$$\parallel$$\mathbf{n}$ and
$\mathbf{e}$$\perp$$\mathbf{n}$.

\begin{figure}[]
 \includegraphics[width=0.45\textwidth]{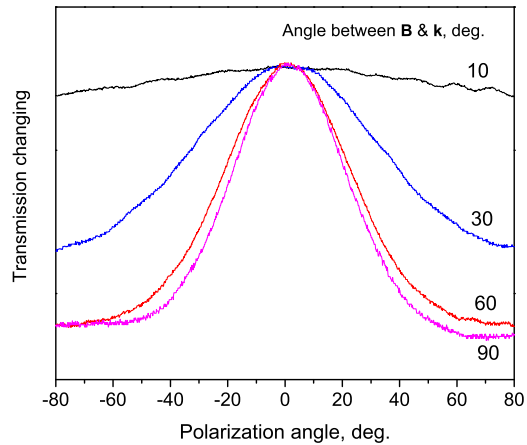}
  \caption{The dependence of the EIT-resonance amplitude $A_{\bf k}$($\varphi$) for the $^{87}$Rb cell transmission on the angle $\varphi$ between $\mathbf{e}$ and $\mathbf{n}$. Angles between $\mathbf{B}$ and $\mathbf{k}$ are shown on the right side of each curve (both vectors lie in the horizontal
plane).}\label{resonance_amplitude}\hfill
\end{figure}

\begin{figure}[t]
  \includegraphics[width=0.47\textwidth]{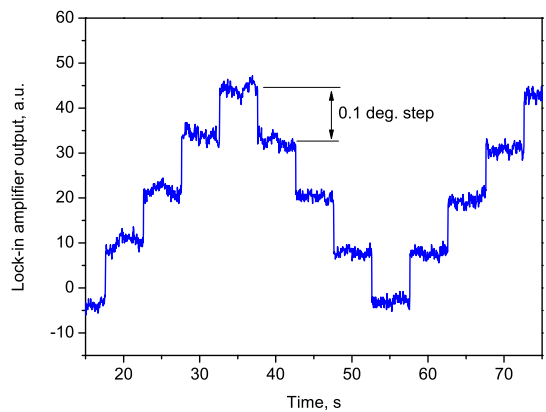}\hfill
  \caption{The lock-in amplifier output at $0.1^\circ$ angle step  variation of the magnetic field direction.  The magnetic field magnitude is 1 G,
  and the angle between magnetic field and wave vector is varied near 90$^\circ$.
  }
  \label{lock-in}
\end{figure}

Before entering the cell, light passes through a half-wave plate,
which is rotating at a 13~Hz rate. As a result, we detect the
dependence of light transmission as a function of the angle
$\varphi$ between  $\mathbf{e}$ and ${\bf n}$, see
Fig.~\ref{resonance_amplitude}. It is worth  noting  that the light
transmission is affected by changes of the EIT transmission and by
variation of the Doppler absorption profile due to optical pumping.
To avoid this distortion of the transmission we detect signals at
the second harmonic of the rf-modulated polarization which is done
by Faraday modulator at 7.6~kHz. To determine the detection
sensitivity of the vector $\mathbf{B}$ direction we change the
orientation of the magnetic field in $0.1^\circ$ steps. The lock-in
amplifier output detects these steps, from which we  estimate a
sensitivity of $\sim$$10^{-3}$~deg$/$Hz$^{1/2}$
(Fig.~\ref{lock-in}). These data were taken for ${\bf B}\perp {\bf k}$
at 1~G magnetic field with the detection bandwidth of 300~Hz.

\begin{figure}[]
  \includegraphics[width=0.47\textwidth]{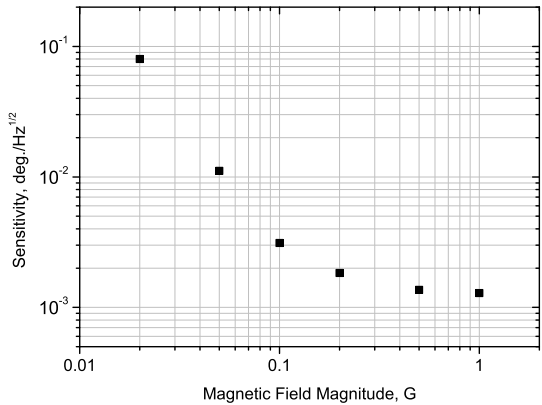}\hfill
  \caption{Compass sensitivity vs. magnetic field magnitude.}
  \label{MF range}
\end{figure}

We have found that the sensitivity depends on the magnitude of the
applied magnetic field (Fig.~\ref{MF range}). At low magnetic field
the sensitivity decreases almost by two orders of value compared to
that at 0.1-7~G. This occurs due to trap states belonging to the
degenerate Zeeman sublevels  of the same hyperfine level where atoms
``hide''. The contrast (as well as signal/noise ratio) grows with
the magnetic field. It is caused by lifting of the sublevel
degeneracy. To destroy  trap states, a magnetic field should be
strong enough, i.e. such that the splitting between the ($0$-$0$)
(i.e. $\Lambda_{1,2}$) and ($1$-$1$) (i.e. $\Lambda_{3,4}$)
transitions greater than the EIT resonance width. Once the ($0$-$0$)
and ($1$-$1$) transitions are separated by $\sim$0.1~G (in our
setup), the compass has the best sensitivity for $|{\bf B}|$$>$1~G.
For some magnetic field ($|{\bf B}|$$>$5~G in our experiments) the
central resonance begins to split \cite{icono05,lin||lin JETP,
Serezha,Kazakov}, because the $\Lambda_1$ and $\Lambda_2$
transitions have a small difference in $g$-factors ($\pm$2.8~kHz/G) due
to the nuclear spin. However, this effect itself does not set the
upper operational limit of the magnetic field for the vector
measurements (compass), because in this case we can work with one of
the two separated $\Lambda$-resonances. We believe that the upper
limitation on the magnetic field in our method is connected with the
degradation of EIT-resonances when the value $\mu_B$$|{\bf B}|$ is
comparable with excited state hyper-fine splitting
$\delta_{hfs}$$\approx$812~MHz, i.e. due to a strong magnetic mixing
between upper hyperfine levels $F_e$=1 and $F_e$=2  (see
Fig.~\ref{levels}). In summary, for the parameters of our setup the
magnetic field operational range of the 3D compass is about
$\sim$0.1-200~G.

\section{Conclusion}

In conclusion, we have developed the generalized principles of
atomic vector magnetometery based on high-contrast EIT-resonances in
a linearly polarized field. These principles follow from a
general symmetry of the problem and are valid for arbitrary atoms, transitions, and arbitrary spectral composition of linearly polarized wave (including a monochromatic wave). The compass involving two non-parallel laser beams
allows to measure the orientation of the magnetic field in three
dimensions. In our proof-of-principle experiment we have achieved a
compass sensitivity $\sim$$10^{-3}$~deg/Hz$^{1/2}$ at intermediate
magnetic fields. We have found that the major contribution to the
noise limiting sensitivity is related to intensity fluctuations of
the laser system. Thus, we believe that the proposed method has a
potential to achieve an angular sensitivity at the level of
$\sim$$10^{-4}$~deg/Hz$^{1/2}$. In contrast to other schemes
of the vector EIT magnetometer, the proposed scheme does not  depend
on a completeness of the magnetometer mathematical model and gives a
straight way to find  the magnetic field direction  and at the end
provides a higher angular accuracy.

We have also discussed  properties and advantages of the EIT
scalar magnetometry, such as  non-sensitivity to quadratic Zeeman
and ac-Stark effects, atom-buffer gas and  atom-atom collisions.
Moreover, our scalar magnetometer works  with a maximal sensitivity
and an accuracy at the arbitrary mutual orientation of the vectors ${\bf
k}$ and ${\bf B}$, i.e. ``dead'' zones are absent (see also \cite{Romalis3}). The spatial resolution, sensitivity, dynamical
range, bandwidth of the magnetic field measurement can be varied by
the proper choice of the cell volume, temperature, buffer gas
type and its pressure (or coating).

EIT vector magnetometers is important for non-invasive
biomedical studies \cite{Bison1,Bison2}, including the temporal and
spacial distribution of the brain and heart electrical currents.
Recent successes in the development of chip-sized atomic clocks and
magnetometers \cite{magnetometry} provide a legitimate optimism for
the creation of a small size magnetic sensor. As a whole, the proposed EIT
compass-magnetometer could find a broad variety of applications in
physics, navigation, geology, biology, medicine, and industry.

We thank L. Hollberg, H. Robinson, J. Kitching, F. Levi, S. Knappe,
V. Shah, V. Gerginov, P. Schwindt, R. Wynands, I. Novikova, E.
Mikhailov, Shura Zibrov  and Yiwen Chu  for helpful discussions.
V.I.Yu. and A.V.T. were supported by RFBR (08-02-01108, 10-02-00591,
10-08-00844) and programs of RAS. V.L.V. and S.A.Z. were supported
by RFBR (09-02-011151).

V. I. Yudin e-mail address: viyudin@mail.ru

S. A. Zibrov e-mail address: serezha.zibrov@gmail.com

\end{document}